\documentstyle[twocolumn,eqsecnum,aps]{revtex}
\def\gtorder{\mathrel{\raise.3ex\hbox{$>$}\mkern-14mu
             \lower0.6ex\hbox{$\sim$}}}
\def\ltorder{\mathrel{\raise.3ex\hbox{$<$}\mkern-14mu
             \lower0.6ex\hbox{$\sim$}}}
\def\msun{\hbox{$\hbox{M}_{\odot}$}}

\begin{document}
\draft
\preprint{CAP--97XXX}
\title{Estimating the detectable rate of capture of stellar mass black holes
by massive central black holes in normal galaxies
}
\author{Steinn Sigurdsson}
\address{
Institute of Astronomy, Madingley Road, Cambridge CB3 0HA, UK\\
}
\date{\today}
\maketitle
\begin{abstract}
The capture and subsequent inspiral of stellar mass black holes 
on eccentric orbits
by central massive black holes, is one of the more interesting
likely sources of gravitational radiation detectable by
LISA. We estimate the rate of observable events and the
associated uncertainties. A moderately favourable mass function
could provide many detectable bursts each year, and a detection
of at least one burst per year is very likely given our
current understanding of the populations in cores of normal 
spiral galaxies.
\end{abstract}

%\pacs{Valid PACS appear here.
%{\tt$\backslash$\string pacs\{\}} should always be input,
%even if empty.}

\narrowtext
\section{
}
\label{sec:level1}

The proposed LISA observatory (\cite{li1,ho95,da93} and this
volume) has optimal sensitivity
to gravitational radiation in the $10^{-3}$--$10^{-2}$ Hz range.
As such it is well matched to the expected frequency of gravitational
radiation emitted by compact objects in the last stages of spiral--in
to massive black holes with total masses of
$M_{BH} \sim 10^6 \, \msun$ (see eg. \cite{th95}).
Previous papers \cite{hi95,sr96} considered the ``likely''
signal from degenerate compact objects in the cusps
of normal galaxies, captured by central black holes such
as are inferred to be present in the Milky Way \cite{ec96} and the nearby
dwarf elliptical, M32 \cite{be96,vm96,vm96b}.

Here we consider the rate for capture of stellar mass black
holes expected to be present in the cusps of normal galaxies,
and estimate the detectable rate of events by LISA, under both
very conservative assumptions and with more optimistic estimates
of the black hole population (see also \cite{sh94,hi96}).

There are several large, systematic uncertainties in estimating
the rate of black hole capture. The proportion of galaxies
harbouring a central black hole is poorly constrained, although
theoretical prejudices suggest central black holes may be
ubiquitous \cite{tr95,ha93,re90}. The mass function of any central black
holes is highly uncertain; there are strong observational
biases on detection of central black holes, and few galaxies
for which there are strong observational constraints as yet
\cite{ko95}. The mass function and total number of stellar mass
black holes is also highly uncertain \cite{ti96}. Stellar mass
black holes may form with masses as low as $2 \, \msun$ and
go up to masses of $50$--$100\, \msun$. The range of masses
of main sequence stars that form a black hole remnant and the
resultant mass of the remnant is also uncertain, and binarity
and metallicity may both affect the remnant mass.
There are observational
biases against detecting the higher mass range of stellar mass
black holes as they may sustain higher accretion rates while
in binaries, and be detectable in our galaxy 
for correspondingly shorter time.

We assume, conservatively, that masses of central black holes
in galaxies
are proportional to the mass of the luminous spheroid of the host
galaxy \cite{tr95},
with a space density of $\sim 3\times 10^{-3}\, {\rm Mpc^{-3}}$
for the range of black hole masses of interest;
and that stellar mass black holes are formed from main sequence
stars with zero--age masses greater than $15\, \msun$, drawn
from a Salpeter mass function extending to $30\, \msun$, with
a total number fraction of $\sim 10^{-4}$ black holes per formed star.
It is possible, if not likely, that this underestimates substantially the
stellar mass black hole fraction.

In the presence of a central black hole, the stellar population
assumes a self--consistent cusp profile, with the stellar
density increasing like $r^{-3/2 -p(r)}$, where $p(r)$ depends
on the local relaxation time and the original phase space
density of the stars \cite{pe72,yo80,sh85,qu95,sr96}.
For (sub)populations of stars for which $p(r) > 0$,
the time scale for two--body relaxation, $t_r$, decreases
with $r$ (assuming $r > r_{coll}$, the collision radius,
at which two--body
scattering becomes ineffective, for black holes $r_{coll} = r_S$).

In the cusp of stars around the central black hole,
stars undergo diffusion in phase space due to
three main processes: the perturbation of their orbits from
the inhomogeneities of the potential, due simply to the finite
number of stars contributing to the mass; large fractional changes
in energy or angular momentum due to elastic scattering by
close encounters with other stars; and, inelastic mergers
with other (extended) stars. Due to the presence of an event
horizon at $r_S$, diffusion in phase space is not symmetric,
rather there is a sink at small radii, and stars with angular
momentum below some critical value are captured by
the central black hole (see \cite{fr76,sr96} for discussion).
The critical region of phase space for which compact stars
are captured is well characterised by a loss--cone
parametrised by a critical angle $\theta_c(r)$;
here we consider the rate for stellar mass black holes in the cusp
to enter $\theta < \theta_c$ through large angle scattering
(ie $\delta \theta \gtorder \theta_c(r)$). The derivation is
discussed in \cite{fr76,sr96}.

The rate of scattering, $R$, into the loss--cone from an
initial orbit of period $P$, is given by \cite{sr96}
\begin{equation}
R(r) = { {N_*^2(r) \theta_c^2(r)}\over {P} } 
\biggl ( {{m_*}\over {M_{BH}} } \biggr )^2 .
\end{equation}
Where $N_*$ is the number of scatterers interior to $r$.
For a given $p(r)$ we can solve for the profile of each stellar component
of the cusp and hence find the rate for capture. Typically the total
rate is peaked at some radius comparable to or larger than $r_c$,
where $\theta_c(r_c) = 1$, unless $r_{coll} \gg r_c$. It is possible
to do a more sophisticated analysis of angular momentum diffusion in
the cusps, but the rate estimates are dominated by uncertainties in
cusp parameters, composition of the stellar population and the black
hole mass, not the details of the scattering.

Stellar mass black holes may be assumed to trace the light into
the cusp, in proportion to their global number fraction. Inside
the main--sequence collision radius, the black hole fraction
increases, as relaxation remains effective for the compact stars
at small radii. For the stellar mass black holes in normal
nucleated spirals, $p(r) \approx 0.7$
at $r \ltorder 10^6 r_S$ but may flatten to $0.3$ inside
$10^4 r_S$.
Neglecting loss by capture, we expect high {\it volume densities} 
of black holes inside $0.001$ pc, even though the total number of black holes
is small.

The inferred rate of black hole capture from such a cusp
is $\gtorder 10^{-5}\ {\rm y^{-1}}$. By assumption there are only
a few hundred black holes in the cusp, and such a rate is not
sustainable. Since the relaxation time for the black
hole population increases with $r$, we expect the inner
cusp to be rapidly depleted of $> 90$\% of its low mass
black holes, with the remainder establishing a steady state
profile with capture rate of order $10^{-8}\ {\rm y^{-1}}$.
Typically the black hole density profile then breaks at
$\gtorder 0.1 \, {\rm pc}$, with the local black hole relaxation time
$\gtorder 10^9$ years. This can provide a sustainable dribbling
of stellar mass black holes from the outer cusp to the
inner cusp, with capture dominated by scattering from
$\sim 10^{3.5} r_S$.

LISA strain sensitivity is $\sim 10^{-23}$ between $10^{-3}$ and $10^{-2}$ Hz.
The characteristic amplitude from a low mass particle, $m_*$
of period $P$ is
\begin{equation}
h_c \sim 4\times 10^{-24} { {m_*/\msun}\over {d/{\rm Gpc}} } 
(M_{BH}/10^6\msun)^{2/3} (10^4 s/P)^{2/3}
\end{equation}
For $m_* \sim 5-10 \,\msun$,
assuming a space density of cuspy galaxies with $M_{BH} \sim 10^6\, \msun$
of $0.003\, {\rm Mpc^{-3}}$ as before, we conservatively expect to see about
three captures per year out to 3 Gpc, if stellar mass black holes are
present at all in cusps of normal nucleated galaxies.

The typical initial eccentricity of a stellar mass black hole
committed to capture by the central black hole is $e_i \sim 0.9995$.
By construction, we have considered the rate for orbits
such that dynamical perturbations will not cause large fractional
changes in the angular momentum of the star, after it is scattered
into the loss--cone, but before it enters the
central black hole. The lifetime to capture through gravitational
radiation emission is well approximated by
$\tau_{GW} = 10^5 (1 - e^2)^{7/2} (P/10^4 s)^{8/3}\, {\rm y}$, or
$\sim 10^6$years for a stellar mass black hole with $e_i=0.9995$ and
apocentre at $O(10^3 r_S)$. 
It is necessary that the period shrink to less than $10^4$ seconds
for the system to be detectable by LISA. As the orbit approaches
$r_S$, and the system departs from the classical regime, the eccentricity
is not well defined (see eg. \cite{sh94}). Detailed evolution
of the orbit is complicated, but we can consider
the qualitative behaviour of the orbit: assuming the peribothron
$r_p = (1 - e) a$ for classical semi--major axis $a$ and some
suitably defined semi--classical eccentricity $e$. Until late stages
of the capture, the orbit decay is well described by the post--Newtonian
quadrupole decay formula, $\dot a = -64 f(e) M_{BH}^2 m_*/5a^3$.
Then $\dot r_p = (1-e)\dot a - a\dot e$. The first term is negative
and much smaller than $\dot a$, while for the orbits we are
interested in $\dot e < 0$ \cite{cf94}, and thus the second term is positive.
Hence evolution of peribothron will not lead to premature
crossing of the event horizon by the star.

In general both the central
black hole, and the stellar mass black holes will be rotating
and the spins and orbital angular momentum of the system will be
uncorrelated. Given high enough signal--to--noise, either from
a system with $m_* \sim 5 \,\msun$ and $d < 1\, {\rm Gpc}$, or
$m_* \sim 50 \,\msun $ at a redshift of $\gtorder 1$, post$^{1.5}$
and higher order spin--orbit couplings may be detected in
the frequency change of the orbit and the orbit precession
with comparable contributions from both the stellar mass
black hole spin and the central black hole spin
(see eg. \cite{cu94}),
enabling measurements of general relativistic effects in the
strong field regime, and correspondingly a unique test of the
validity of general relativity as a theory of gravity in this
regime. In addition, any such detection would provide important
information about the masses and spin of both the central massive
black hole, and the stellar mass black holes present in the
centres of galaxies.

We have assumed that the proportion of stellar mass black holes
in the high density cores of galaxies is comparable to that
estimated for the total stellar population. If there is a systematic
difference in initial mass function, theoretical considerations
suggest the mass function should be biased towards more massive stars,
and hence more massive remnants, in high density environments \cite{la71,la92}.
An initial higher fraction of black holes would lead to more rapid
merger (commencing either a few million years after the first burst
of star formation, if the central black hole formed before the stars;
or, commencing after the central black hole came to dominate the
dynamics of the core, if the central black hole arrived or formed
after the initial stars formed), but the same sustainability
problems exist independent of the number of black holes, provided
there are fewer than $10^4$ initial black holes, and more than $30$
or so. Hence most of the stellar black holes are swallowed in a short time
compared to the lifetime of the galaxy.
There are two factors, however, which lead to a more favourable rate
of bursts: with more initial stellar mass black holes, we expect
more higher ($\sim 30-100 \,\msun$) mass ones, and the capture of those
is detectable by LISA anywhere in the universe; it also possible
that high mass, low metallicity, stars are more
likely to form  massive black hole remnants. If there was even one $50 \,\msun$
black hole in the core of each galaxy now containing a $\sim 10^6 \,\msun$
central black hole, we expect several mergers per year at $z > 1$,
all detectable by LISA. The redshifting of the gravitational radiation
is not likely a concern, as central black holes with masses now $> 10^6 \,\msun$
must have had somewhat smaller masses in the past.

In our own galaxy, there is now substantial circumstantial evidence
that there was a recent starburst in the central region \cite{kr96}.
If such behaviour is typical of normal nucleated spirals, then every
$\sim 10^8$ years, the cusp population of stellar mass black holes
may be partially replenished. If this occurs commonly, rates of
capture in the nearby universe may be an order of magnitude greater
than we have estimated, and we may expect a strong detectable signal 
every few weeks.

Other favourable scenarios for providing detectable signals
include the merger of two massive black hole of comparable
mass, or a massive black hole and an ``intermediate mass'' black hole.
It is possible that even more massive black holes form
in galaxies ($M_{BH} \approx 10^2 - 10^3 \,\msun$), either
during early stages of galaxy formation, as part of the
ultra--low metallicity Population III stars, or in
clusters of stars, growing from $< 10^2 \,\msun $ through
mergers and accretion. If such black holes can reach the inner
parsec of their host galaxy, relaxation will bring them to
small radii and capture by the central black hole. The resultant
burst of gravitational radiation would be easily detectable, but
both the number of such objects and the rate at which they
might reach centres of galaxies is highly uncertain
(see eg. \cite{ha94,po94}).

It seems quite probable that LISA will detect signals from the capture
of stellar mass compact objects by central massive black holes
in normal galaxies, if our current understanding of the properties
of the central stellar population is even marginally correct, and
if the masses of the central black holes in nearby galaxies are at
all typical. It is possible that the rate of detectable events
is greater than one per month. 
The actual detection of such events, in addition to
testing theories of gravity in the strong field regime,
would provide the only data on the mass function of remnants stars in
galactic centres, and would probe the low mass end of the range of
massive central black holes in the distant universe.

\acknowledgments

I would like to thank M. Rees, P. Bender and A.G. Polnarev
for helpful discussions.
The author gratefully acknowledges
an EU DGXII TMR Category 30 Marie Curie personal Fellowship.

\end{document}